\documentclass[8.5pt,twoside,twocolumn]{article}
\usepackage[super,sort&compress,comma]{natbib} 
\usepackage{mhchem}
\usepackage{times}
\usepackage{sectsty}
\usepackage{balance} 
\usepackage{amsmath}

\usepackage{graphicx} 
\usepackage{lastpage}
\usepackage[format=plain,justification=raggedright,singlelinecheck=false,font=small,labelfont=bf,labelsep=space]{caption}
\usepackage{fancyhdr} 
\usepackage{color}

\def\dee{\mathrm{d}} 
\def\kT{\ensuremath{k_{\mathrm{B}} T}} 
\def\vec#1{{\mathbf{#1}}} 
\def\vr{\vec r} 
\def\vR{\vec R} 

\begin{document}

\thispagestyle{plain}
\fancypagestyle{plain}{
\renewcommand{\headrulewidth}{1pt}}
\renewcommand{\thefootnote}{\fnsymbol{footnote}}
\renewcommand\footnoterule{\vspace*{1pt}%
\hrule width 3.4in height 0.4pt \vspace*{5pt}} 
\setcounter{secnumdepth}{5}

\makeatletter
\def\subsubsection{\@startsection{subsubsection}{3}{10pt}{-1.25ex plus
    -1ex minus -.1ex}{0ex plus 0ex}{\normalsize\bf}}
\def\paragraph{\@startsection{paragraph}{4}{10pt}{-1.25ex plus -1ex
    minus -.1ex}{0ex plus 0ex}{\normalsize\textit}}
\renewcommand\@biblabel[1]{#1} \renewcommand\@makefntext[1]%
                       {\noindent\makebox[0pt][r]{\@thefnmark\,}#1}
                       \makeatother
                       \renewcommand{\figurename}{\small{Fig.}~}
                       \sectionfont{\large}
                       \subsectionfont{\normalsize}

\fancyfoot{}
\fancyfoot[RO]{\footnotesize{\sffamily{1--\pageref{LastPage} ~\textbar  \hspace{2pt}\thepage}}}
\fancyfoot[LE]{\footnotesize{\sffamily{\thepage~\textbar\hspace{3.45cm} 1--\pageref{LastPage}}}}
\fancyhead{}
\renewcommand{\headrulewidth}{1pt} 
\renewcommand{\footrulewidth}{1pt}
\setlength{\arrayrulewidth}{1pt}
\setlength{\columnsep}{6.5mm}
\setlength\bibsep{1pt}

\twocolumn[
  \begin{@twocolumnfalse}
\noindent\LARGE{\textbf{Designing stimulus-sensitive colloidal walkers}}
\vspace{0.6cm}

\noindent\large{\textbf{Francisco
    J. Martinez-Veracoechea,\textit{$^{a}$} Bortolo
    M. Mognetti,\textit{$^{a,b}$} Stefano
    Angioletti-Uberti,\textit{$^{a,c}$} Patrick
    Varilly,\textit{$^{a}$} Daan Frenkel,\textit{$^{a}$} and Jure
    Dobnikar$^{\ast}$\textit{$^{a,d}$}}}\vspace{0.5cm}

\noindent\textit{\small{\textbf{Received Xth XXXXXXXXXX 20XX, Accepted
      Xth XXXXXXXXX 20XX\newline First published on the web Xth
      XXXXXXXXXX 200X}}}

\noindent \textbf{\small{DOI: 10.1039/b000000x}}
\vspace{0.6cm}

\noindent \normalsize{Colloidal particles with DNA ``legs'' that can
  bind reversibly to receptors on a surface can be made to `walk' if
  there is a gradient in receptor concentration. We use a combination
  of theory and Monte Carlo simulations to explore how controllable
  parameters, e.g. coating density and binding strength, affect the
  dynamics of such colloids. We find that competition between
  thermodynamic and kinetic trends imply that there is an optimal
  value for both, the binding strength and the number of ``legs'' for
  which transport is fastest. Using available thermodynamic data on
  DNA binding, we indicate how directionally reversible,
  temperature-controlled transport of colloidal walkers can be
  achieved. In particular, the present results should make it possible
  to design a chromatographic technique that can be used to separate
  colloids with different DNA functionalization.}
\vspace{0.5cm}
 \end{@twocolumnfalse} ] 
 
 
\section{Introduction}

\footnotetext{\textit{$^{a}$~University of Cambridge,The University
    Chemical Laboratory, Lensfield Road, CB2 1EW, Cambridge, UK}}
\footnotetext{\textit{$^{b}$}~Center  for Nonlinear Phenomena and Complex Systems, 
Universit\'e Libre de Bruxelles, Code Postal 231, Campus Plaine, B-1050 Brussels, Belgium}
\footnotetext{\textit{$^{c}$}~Department of Physics,
  Humboldt-University Berlin, Newtonstrasse 15, 12489 Berlin, Germany}
\footnotetext{\textit{$^{d}$}~Department of Theoretical Physics, Jo\v
  zef Stefan Institute, Jamova 39, 1000 Ljubljana, Slovenia}
\footnotetext{\textit{$^{\ast}$}~jd489@cam.ac.uk}

The study of DNA-coated colloids (DNACCs) started with two seminal
papers , one by the group of Mirkin~\cite{mirkin} and one by
Alivisatos and collaborators~\cite{alivisatos}. Since then, DNACCs
have become a vibrant area of research that includes experiments
\cite{mirkin,alivisatos,
  mirkin-crystal,gang-crystal,crocker-tweezers,rogers-crocker,control-position,mirjam-nature,mirjam-jacs,
  dreyfus,lorenzo-bigels, mirkin-crystal2, gang-crystal2,
  lorenzo-kinetics}, and modelling~\cite{stefano-nature, tkachenko,
  bortolo-pnas, patrick-jcp, stefano-jcp, bortolo-competing, olvera,
  travesset, Fran2, bianca}.  Recent reviews covering both aspects
DNACC-based materials science can be found in
Refs.~\cite{lorenzo-review} and \cite{travesset-review}. The goal of
most studies to date was to induce the self-assembly of predesigned
crystalline structure of DNACCs, by tuning the DNA-mediated colloidal
interactions~\cite{mirkin-crystal, gang-crystal, stefano-nature,
  gang-crystal2, mirkin-crystal2}. Yet, other promising areas where
DNA programmability can be exploited are quickly emerging: a notable
example is development of DNA-based
motors~\cite{turberfield-walker,seeman-walker} that move in a
programmable and reproducible way. Such motors could find applications
in the step-by-step synthesis of macromolecules
\cite{turberfield-synthesis,seeman2,liu}, or in DNA-based computing
\cite{dna-computing1,dna-computing2}.

Here we investigate how DNACCs functionalized with different
nucleotide sequences can be used as nano- or micro-scale
transporters. We study colloids coated with DNA molecules that are
mostly composed of inert double-stranded DNA (dsDNA) sequences but 
have a `sticky' single-stranded DNA (ssDNA) end. The colloids
move on surfaces coated with concentration gradients of ssDNA strands
with complementary sequences. Using a combination of theory
\cite{patrick-jcp,stefano-jcp} and Monte Carlo (MC) simulations, we
analyse how such directional motion depends on experimentally tunable
parameters, such as the number of DNA ``legs'' on each colloid, the
hybridisation free-energy of their sticky ends, and the grafting
density of the surface strands. We demonstrate that it is possible to
engineer a system such that the motion can be reversed by changing
temperature, pH, salt concentration, or other external
parameters. Such switchability would enable novel applications like
cyclable transport of specific substances on DNA tracks. Finally, we
indicate how our results could be used to design a chromatographic
tool to separate DNACCs with different functionalization.
\begin{figure}
\centering 
\includegraphics[width=0.235\textwidth]{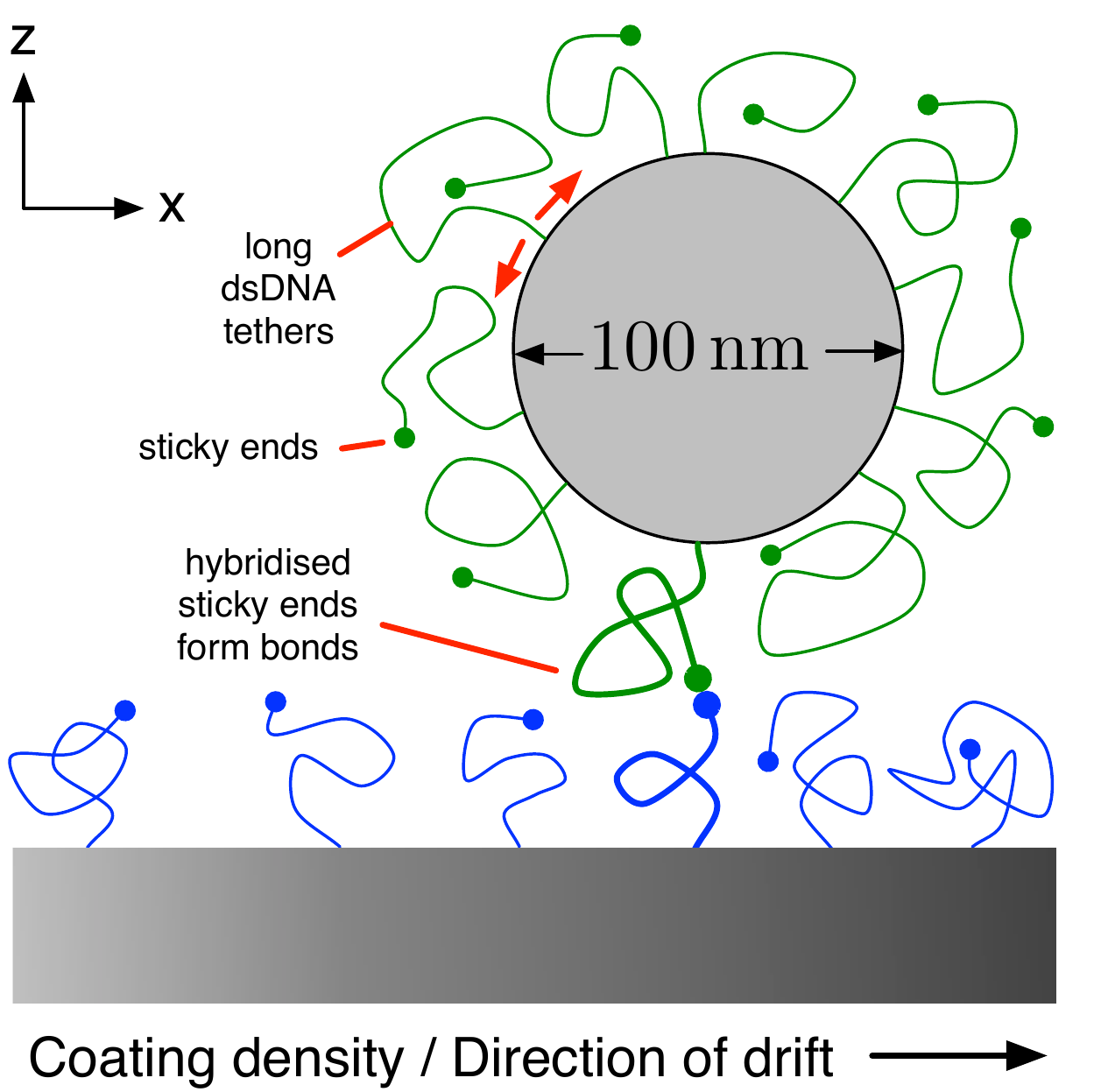}
\includegraphics[width=0.235\textwidth]{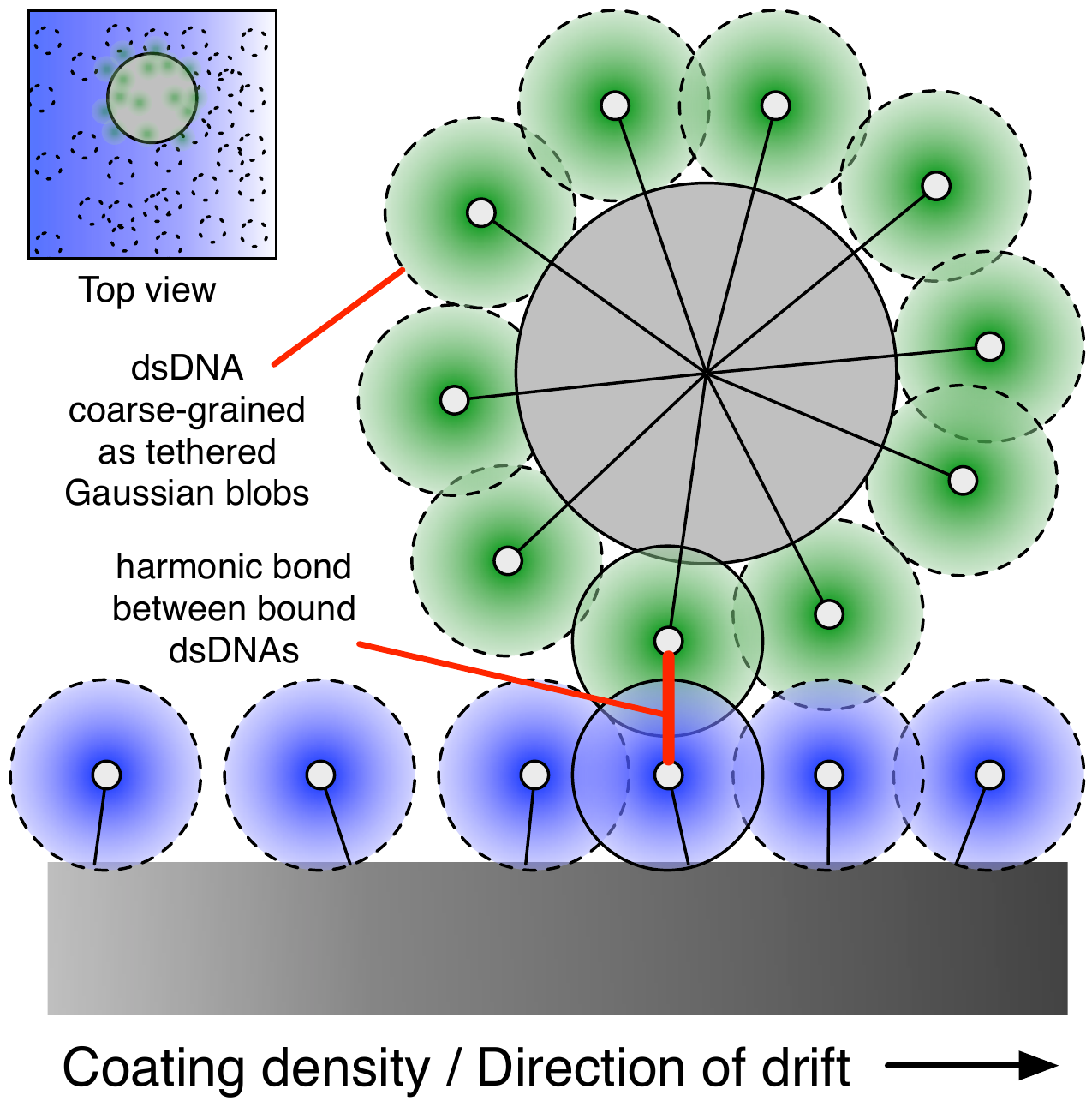} 
\caption{\label{fig:nano_colloids} Nanocolloids walking on a
  DNA-coated surface. Atomistic detail (left) and coarse-grained
  representation (right).}
\end{figure} 
 
\section{Modelling of DNA-based walkers} 
In Figure~\ref{fig:nano_colloids} we show a typical setup for DNACCs
walking on a coated surface. To model DNACCs we use a coarse-grained
model~\cite{Fran1, Fran2} that has been shown to reproduce all
features of the phase behaviour of DNACCs
(Figure~\ref{fig:nano_colloids}b) (for details of the model please see the 
Methods section). In this model, the colloids are
represented as hard-spheres with radius $R_c$. We assume that the DNA
chains are long enough to be modelled as soft
``blobs''~\cite{BarbaraBlobs}, whose only degree of freedom is the
position of the center of mass. Blobs interact with each other through
a Gaussian repulsion and with colloids or the flat surface via an
exponential repulsion. To each colloid, we attach a number~$\kappa$ of
DNA chains. A harmonic spring represent the binding of the DNA chain
to the surface.  These attached blobs act as ``legs'' that can freely
move along the surface of the colloids~\cite{Fran1}. The surface $z=0$
is grafted with immobile DNA blobs that are distributed in a random
way along the $y$ axis and with a linear grafting density gradient
along the~$x$ axis. In all cases discussed below, we average all
observations over multiple realisations of the surface grafting
points.

DNAs on the colloids have a single-stranded sticky end that is
complementary to the sticky ends on the surface. A given pair of
complementary DNAs is either bound or unbound. If two DNA blobs are
bound, the free energy of the system shifts by an amount~$\beta f$
($\beta =1/k_B T$, where $k_B$ is Boltzmann's constant and $T$ the
temperature), and the centers of mass of the bound blobs interact via
an additional harmonic spring potential \cite{Fran1}. The binding
energy $\beta f$ is related by a constant shift to the hybridisation
free energy of the sticky ends in solution at standard conditions, for
colloids with diameter $100\,$nm, we have $f \approx \Delta G_0 +
11.6\,\kT$ (see Materials and Methods section for more detail), where
$\Delta G_0$ is the hybridisation free energy of complementary
single-stranded sticky ends forming a bond.

On a surface with receptor gradient, the DNACCs are subject to an
effective force in the direction of the gradient. The interaction
free-energy between colloid and substrate is determined by a
combination of the hybridisation free energies of the single bonds and
by the combinatorial entropy due to the formation of multiple bonds
between the particles and the surface. The free energy profile can be
evaluated using a recently developed self-consistent mean-field
theory~\cite{patrick-jcp,stefano-jcp}.

The mean-field theory (described in detail in the Appendix) disregards
DNA-DNA excluded volume interactions and is therefore approximate (but
qualitatively correct) for nano-sized colloids that typically have
relatively high coating densities. For micron-sized colloids, which
display lower functionalization densities, the neglected interactions
are almost negligible, hence the theory approaches quantitative
accuracy ~\cite{patrick-jcp,stefano-jcp} and can be used as a powerful
tool to explore the mechanisms of colloidal motility.

The calculated free energy profiles (see
Fig. \ref{fig:MicronWolloids1} in the Methods section) feature two
regimes characterized by the dominance of either the enthalpy of
hybridisation or the combinatorial entropy. As expected, the DNACCs
generally experience a strong driving force to ``walk'' up the
concentration gradient. This thermodynamic driving force monotonically
increases as the number of DNAs per colloid $\kappa$ and the binding
strength ($\beta f$ more negative) are increased.

However, the free-energy gradients do not completely determine the
efficiency of colloidal transport on the surfaces: if the
hybridisation free energy $\beta f$ becomes too negative, the kinetics
of unbinding slows down. Similarly, when the number of ``legs''
$\kappa$ is increased, it becomes more difficult for the colloid to
move as it is anchored in multiple places. To explore the competition
between the opposing thermodynamic and kinetic effects we performed
kinetic MC simulations where we initially put the colloids at $x=0$
and let them diffuse freely. Details of the kinetic Monte Carlo moves
are described in the Materials and Methods section. For the trial
moves that account for bond-forming/bond-breaking we followed the
procedure described in ref.~\cite{Fran1}. We kept track of the
position of the colloid at different simulation times. Since we are
interested in the regime where the motion of the colloid is entirely
due to DNA hybridization, we impose the restriction that the
translation moves of the colloid are only attempted if at least one of
its DNA chains is hybridized.
\begin{figure} 
\centerline{\includegraphics[width=0.5\textwidth]{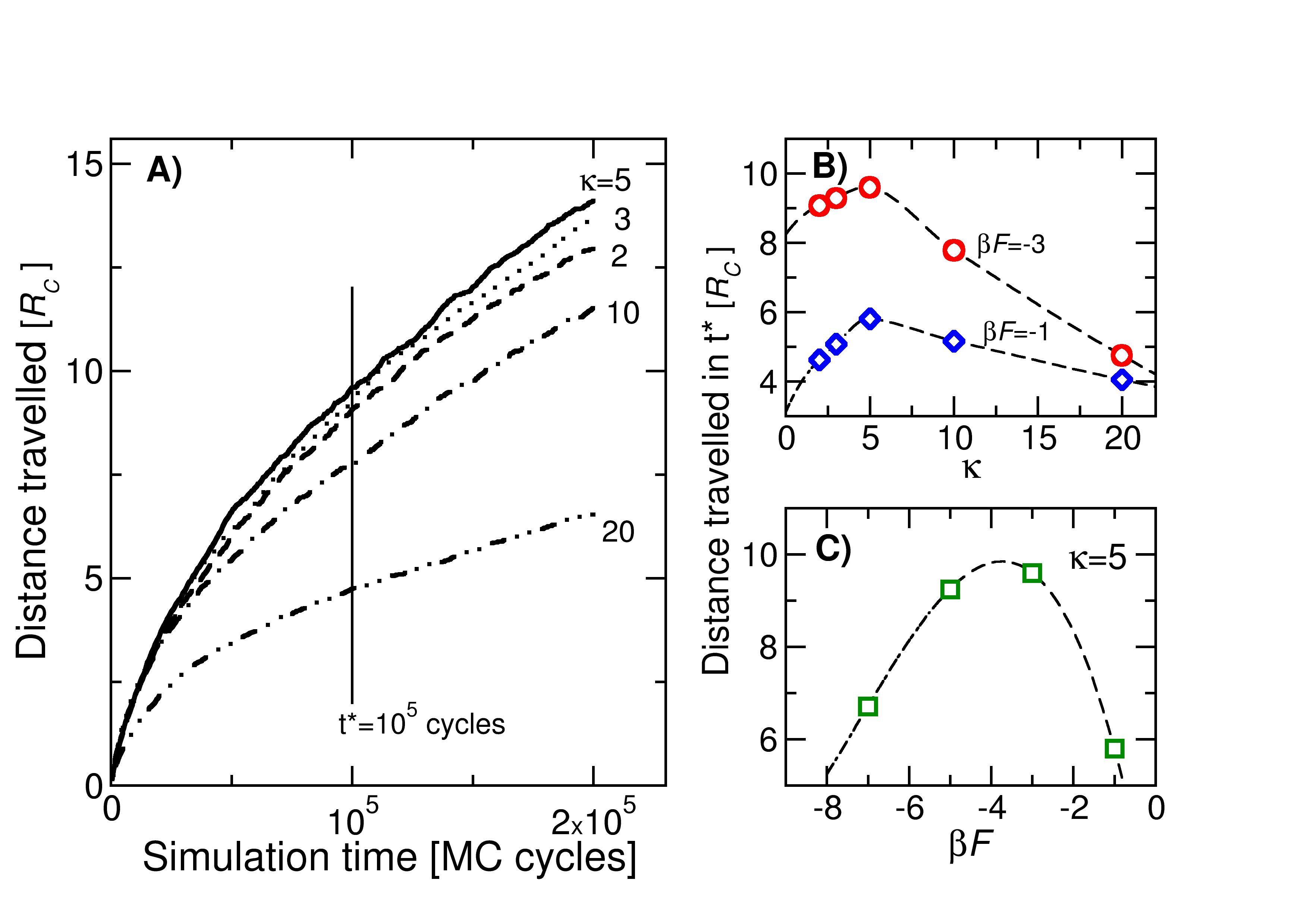}} 
\caption{\label{fig:FranWalk} Walking efficiency of colloidal walkers
  in kinetic MC simulations on surfaces with linear gradient of the
  receptor density. {\bf A)}: Distance  travelled  as a
  function of the simulation time, for colloids with different number
  of ``legs'' $\kappa$ from 5 (top curve) to 20 (bottom curve). The
  value of the binding strength is fixed to $\beta F=3$. The distance
  travelled in $t^*$ cycles as a function of $\kappa$ at fixed $\beta
  F$ ({\bf B))}, and as a function of the binding strength $\beta F$
  at fixed $\kappa=5$ ({\bf C)}). }
\end{figure}  

In Fig. \ref{fig:FranWalk}{\bf (a)} we show the average of the
distance travelled by colloids with different values of $\kappa$ at
$\beta f=-3$. The averaging was based on 2000 trajectories.  On
average, the colloids move in the direction of increasing
concentration of surface DNA as time progresses. The kinetics of the
system is governed by the rate of hybridisation and breaking of the
single bonds. In~\cite{Crocker_Tau} the lifetime of a single bond has
been shown to be of the order of $\mu s$. Since we typically attempt
about 10 to 100 binding/unbinding events in a single MC cycle, the
corresponding real timescale can be estimated. A rough estimate
predicts that the walking speeds can be up to microns per second,
which is comparable to the speed of biological molecular motors.  Of
course, the actual speed depends on the design parameters of the
system. {Our kinetic simulations reveal (Fig. \ref{fig:FranWalk} {\bf
  (b)}, {\bf (c)}) that an optimum value of $\beta f$ and $\kappa$
exist where colloids walk the fastest in the direction of the
concentration gradient: $\kappa_{opt}\approx 5$ and 
$\beta F_{opt} \approx-4$. While these exact values should depend 
to some extent on the model and simulation details, the existence of an 
optimal parameter set $(\kappa_{opt},\beta F_{opt})$ is due to 
the competition between kinetics and thermodynamics, 
and is thus a general feature of the system.}

The observed behaviour suggests a chromatographic method to separate
DNACCs with different functionalizations. In a traditional
chromatographic column each reaction product moves over a different
distance owing to a different interaction with the substrate, and the
same happens for DNACCs that differ in the number of DNA
``legs''. Especially for the case of nano-sized colloids where the
average number of strands show greater relative variation due to their
small size, obtaining uniformly functionalized colloids is important
for applications. For instance, statistical fluctuations of the number
of grafted DNA may interfere with the formation of ordered
crystals. We suggest a simple method to characterize the colloids: on
a coated surface with a gradient in receptor concentration the
colloids coated with complementary sticky ends will move along the
gradient. At a given temperature, their speed will depend on the
functionalization $\kappa$ and after a finite time they will be
spatially separated. Since, due to the non-monotonic behavior of
walking efficiency (Fig.~\ref{fig:FranWalk}), the separation will not
be complete, we could repeat the process at a different temperature in
order to further differentiate the colloids. Our results allow us to
design -- according to the experimental details of the system -- an
optimal temperature cycle for efficient separation.

\section{Designing reversible DNACCs walkers}  
\begin{figure}[h] 
\vspace{1cm} 
\centerline{\includegraphics[width=0.5\textwidth]{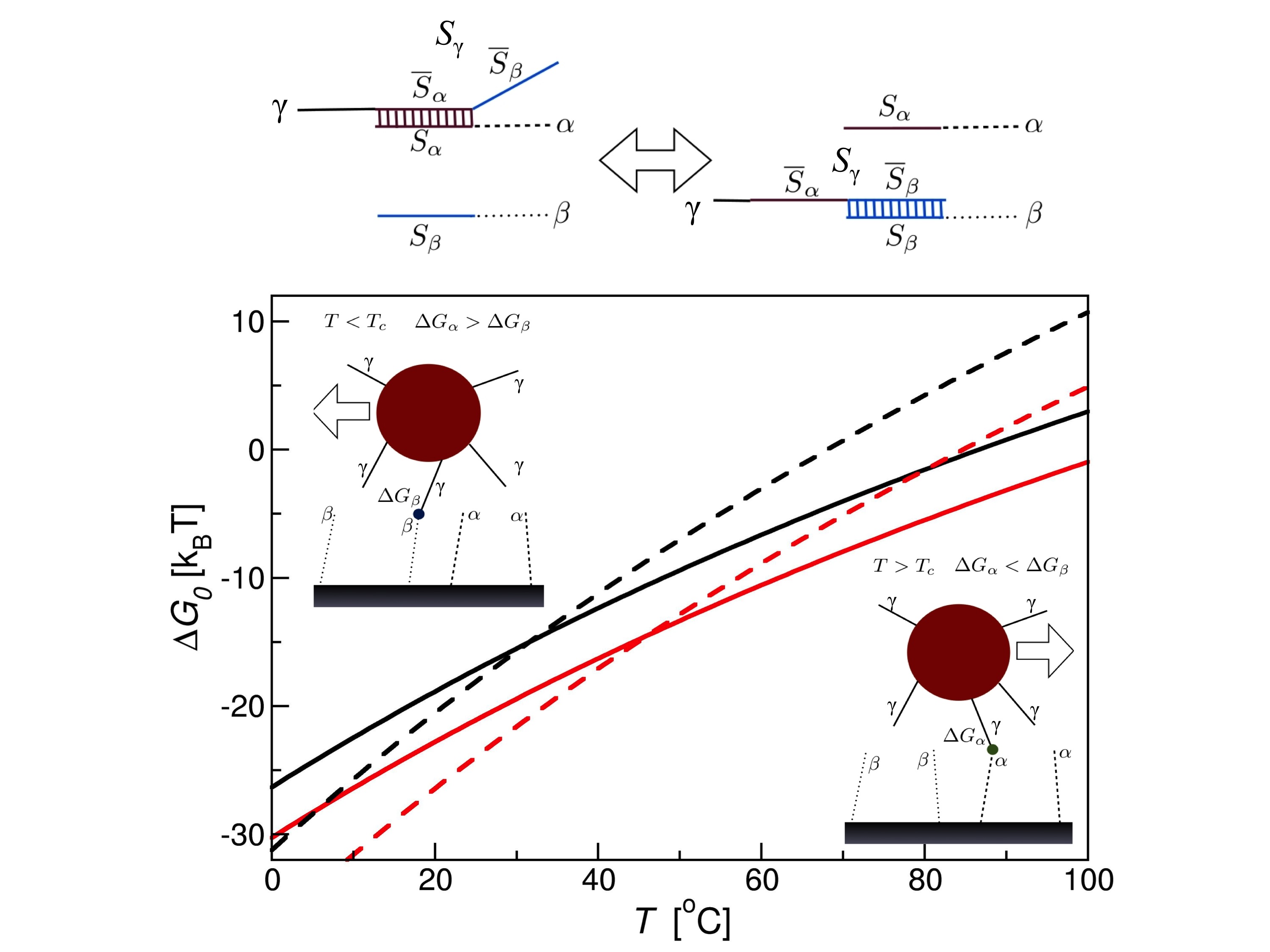} } 
\caption{Sticky-end architecture. The sticky ends on the colloidal
  ``legs'' are designed so that they can bind to two different
  receptors $\alpha$ and $\beta$. The relative binding strength and
  thus the direction of motion can be regulated by temperature as
  shown in the plot of the binding free energy as a function of the
  temperature. The calculations were made for the strands in  
  Eq.~\ref{Sticky1}; the data shown is for sticky ends $S_\gamma$ 
  binding to the substrate receptors $S_{\alpha}$ (solid lines) and 
  $S_{\beta}$ (dashed lines). Results are plotted for two different 
  salt concentrations: 0.06 M (black) and 1M (red). The sequences 
  have been designed so that there is a transition temperature $T_c$, 
  at whicg the preferable binding changes from $\alpha$ ($T<T_c$) 
  to $\beta$ at $T>T_c$.}
\label{FigSys}
\end{figure}

In the system described above, the colloidal motion can be controlled
by external parameters, e.g. temperature, salt concentration or pH.
As a proof of principle, we show here how to design DNACCs walkers
with the ability to reverse their direction of motion as a function of
temperature.  The approach that we propose is based on the possibility
to swap between two kinds of linkages (see Fig.~\ref{FigSys}). This is
encoded in the hybridisation free energies of the two competing
linkages ($\Delta G_\alpha$ and $\Delta G_\beta$) that are designed to
invert their relative strength as the temperature is lowered. In order
to engineer this changeover behaviour, it is useful to decompose the
hybridisation free energy into the configurational term ($\Delta G_{c}
$) and the sticky-ends term ($\Delta G_{x0} $)
\begin{eqnarray} 
\Delta G_x &=& \Delta G_{c} + \Delta G_{x0} \qquad x=\alpha,\beta \;. 
\label{Eq:Gx}
\end{eqnarray} 
$\Delta G_{x0}$ depends on the sticky-end sequences and is highly
sensitive to the temperature, whereas $\Delta G_{c}$ is an entropic
term depending on the accessible configurations for the tethering part
of DNA (i.e., that which is not involved in the hybridisation into
dsDNA when bonds are formed). To a first approximation, $\Delta
G_{c}$, when measured in units of the thermal energy $k_B T$, is
independent of temperature. In principle one could use different
tethering parts to control the binding free energy. However, this
requires detailed calculation of $\Delta G_{c}$, which requires a
detailed model of the tethering polymers and its interactions.  Such a
model may not always be available.  For design purposes it is
therefore better to consider a system of colloids for which all sticky
ends are tethered to the colloidal surface by the same type of
polymer. This implies that the tether has no effect on the relative
strength of different bonds and that the only parameter we 
need to tune is $\Delta G_{x0}$, the hybridisation free energy of the
bonding pairs. We recall that $\Delta G_{x0}$ is the sum of
enthalpic and entropic contributions
\begin{eqnarray} 
\Delta G_0 &=& \Delta H_0 - T \Delta S_0 \;, 
\label{Eq:G0} 
\end{eqnarray} 
where -- to a good approximation -- $\Delta H_0$ and $\Delta S_0$ only
depend on the salt concentrations and the nucleotide sequences. Both
terms can be calculated reliably by the nearest neighbor rules of
SantaLucia \cite{santalucia} that express these quantities for every
possible pair of strands as a sum over twelve possible oriented
nearest neighbour.  Since the entropic term in the free energy
$T\Delta S$ increases linearly with temperature, whereas the enthalpic
term remains roughly constant, the hybridisation free-energy will vary
linearly with temperature.

{ 
In view of the above, we have constructed 
the sequences $s$ as a series of non-reactive poly-T strands 
($T_n$) that connect the tethering points with the reactive 
sticky ends $S$: 
\begin{eqnarray}
s_{x} = 5' - (T_n)\,S_x - 3'\;,
\label{Eq:s}
\end{eqnarray}
where index $x=\gamma,\alpha,\beta$ is used to distinguish between the 
colloidal "legs" $s_\gamma$, and the two types of surface 
receptors $s_{\alpha,\beta}$.\footnote{Note that the length of the ($T_n$) 
sequences contributes to the configurational part of the 
hybridisation free energy, $\Delta G_{c}$, which 
-- as argued above -- does not depend on temperature.} 
The notation 5' and  3' is the standard way to label the directionality 
of DNA sequences that affects the hybridisation free energy of the sticky ends,  
$\Delta G_{x0}$. 

In order to design appropriate sticky ends $S_x$, we considered the case 
where the sticky end of the colloidal ``leg'' $S_\gamma$ consists of two 
nucleotide sequences, $S_\gamma=\bar{S}_\alpha+\bar{S}_\beta$, and 
can bind to two complementary receptor sequences  $S_\alpha$ and 
$S_\beta$ (see Fig.~\ref{FigSys}). We chose the $S_\alpha$
complementary part to be shorter than $S_\beta$ but with a larger 
number of "strong" $G\!\!-\!\!C$ bonds, as opposed to "weak" $A\!\!-\!\!T$ bonds
\cite{santalucia}.  For such sequences, we expect the following
behavior: at high temperatures, where the entropic term in 
Eq.~\ref{Eq:G0} dominates, the shorter $\mathrm{S_\gamma\!\!-\!\!S_\alpha}$ 
pair will have a lower hybridisation free energy: $\Delta G_\alpha < \Delta G_\beta$. 
As temperature decreases, the entropic penalty becomes less relevant 
with respect to the enthalpic gain, and the free energy $\Delta G_\beta$ of the longer 
$S_\gamma\!\!-\!\!S_\beta$ pair that is capable of forming more bonds will be lower.
Since $\gamma$ can bind either to $\alpha$ or to $\beta$ but not likely to both 
\footnote{When tethered to a substrate, simultaneous binding of the same DNA
  filament to two other strands would require a high entropic penalty
  $\Delta G_{c}$ and thus cannot occur.  This is not true for free
  strands in solution, a difference that is exploited in the design
  of DNA-origami}, a crossover in binding affinity is expected at a temperature 
$T_C$. By varying the length of the sequences and the relative number of strong
$G-C$ bonds,  the relative hybridisation free energy of the two sequences 
and the crossover temperature $T_C$ can be controlled. We have used the 
DINAMelt web server~\cite{DINAMelt} to evaluate the hybridisation free energy 
(Eq.~\ref{Eq:G0}) and designed the sticky-end sequences in such a way 
as to observe the crossover at temperature between $30^\circ C < T_C < 50^\circ C$. 
Two sticky-end sequence sets $S_x$ that we designed in this way are:
\begin{eqnarray} 
S_{\gamma} &=&\underline{TTGAGAAATCCCCCCCCC}
\nonumber \\ 
S_{\alpha}&=&\underline{GGGGGGG}
\nonumber \\
S_{\beta}&=&\underline{GGATTTCTCAA}
\label{Sticky1} 
\end{eqnarray} 
and 
\begin{eqnarray} 
S_{\gamma} &=&\underline{GGATCAATCTTGGGGGGG} 
\nonumber \\ 
S_{\alpha} &=&\underline{CCCCCCC}
\nonumber \\ 
S_{\beta} &=&\underline{AAGATTGATCC}\;. 
\end{eqnarray} 
The dependence of the hybridisation free energy $\Delta G_0$ for one of 
the above solutions (Eq.~\ref{Sticky1}) on temperature is shown on 
Figure \ref{FigSys}. We see that the binding free energy indeed has a 
crossover temperature, at which the relative binding strength of the 
colloidal legs to receptors $\alpha$ and $\beta$ changes sign. The red 
and black set of curves illustrate the dependence of the crossover 
temperature on salt concentration. }

Using our ability to switch the relative strength of $\alpha$ and
$\beta$ bonds, we can now design a system where the direction of the
walkers' motion can be changed with temperature. We assume that the
surface has been prepared such that its overall grafting density
$\sigma$ is uniform, however the densities of the two strands
$\sigma_\alpha$ and $\sigma_\beta$ have gradients in opposite
directions, e.g.  $\sigma_\alpha=\sigma (1-x/L)$ and
$\sigma_\beta=\sigma x/L$. On such a surface, the colloidal walker
will follow the direction that maximizes the number of strong bonds,
since the combinatorial entropy due to the different possible
configurations of bonds remains constant throughout the
system. However, the relative strength of the bonds depends on
temperature and hence, changing the temperature one can reversibly
change the direction of walking. It is possible to design similar
systems with the pH or salt concentration -- instead of the
temperature -- as the control parameter. We note that many examples of
unidirectional motion of colloids along surface gradients have been
reported in the literature
\cite{tkachenko-pre,turberfield-walker,seeman-walker}. However,
to the best of our knowledge, the technique that we propose to control
the direction of motion of the colloids is new.

{ The driving force moving the reversible walkers is the replacement of weak 
bonds with stronger ones (this regime has been defined `enthalpic' in Sec. 5.3). 
For this reason, directed motion (in contrast with random diffusion) can be 
achieved also when all strands on the colloids are hybridised. In this respect 
the reversible walkers are analogous to a vitrimer system, for which any 
rearrangement is achieved without varying the degree of cross-linking between 
polymers~\cite{vitrimers}. In vitrimers, swapping between different bonding 
configurations can be achieved only in the presence of an appropriate catalyst 
and at sufficiently high temperatures. These requirements stem from the necessity 
to effectively break strong covalent bonds before reforming them with different 
partners. Although DNA-hybridisation for the typical sequences we describe 
has a much lower energy than any covalent bond, the binding strength can still 
be considerable, especially at low temperatures, thus the single-bond breaking 
kinetics can be very slow. Under such conditions, the mobility of the colloid 
(i.e. the proportionality constant between its speed and the free-energy gradient) 
is dominated by this single-bond breaking process~\cite{bortolo-competing}. 
In the field of DNA-nanotechnology some strategies have been devised to 
alleviate this problem, e.g. toe-holding mediated strands 
displacement~\cite{Winfree}. However, a detailed discussion of how these 
schemes could be used in our system goes beyond the scope of the present work.}

The ability to reverse the direction of colloidal motion could be
interesting for a variety of applications.  Moreover, it would allow
us to refine the chromatografic method proposed above: if we coat the
surface with two kinds of receptors $\alpha$ and $\beta$ with the
binding free energies featuring a crossover as in Fig.~\ref{FigSys}
and if the two strands have density gradients in orthogonal
directions, the direction of colloidal motion will depend on the
temperature. By, e.g. gradually changing the temperature around the
crossover temperature, we can drive the colloids along the curved
trajectories, as well as separate them according to their
functionalization. Finally, one could envisage separating colloids
with similar sequences differing in only one or two nucleotides.

\section{Methods}
\subsection{Modeling polymers and bond formation}
We use a previously developed coarse-grained model~\cite{Fran1, Fran2}
to represent the DNACCs (Figure~\ref{fig:nano_colloids}b).  We model
colloids as hard-spheres of radius~$R_c$, and the surface that they
walk on as a hard half-space~$z < 0$.  The bare colloid-colloid
interaction $V_{HS}(d)$, is given by
\begin{equation} 
V_{HS}(d) = \begin{cases} 
\infty,&d < 0;\\ 
0,&d \geq 0, 
\end{cases} 
\end{equation} 
where $d$ is the distance of closest approach between two colloids or
between a colloid and the surface ($d=2R_c$ for two colloids and
$d=R_c$ for colloid-surface).\\

The long, flexible dsDNA chains (gyration radius $R_g = R_c / 3$) are
modelled as soft ``blobs''~\cite{BarbaraBlobs}, whose only degree of
freedom is the position of the center of mass.  Blobs interact with
each other through a Gaussian repulsion, given by
\begin{equation} 
 \beta V_{bb}(r) = 1.75\, e^{-0.80 (r/R_g)^2}\;, 
\end{equation} 
where $r$ is the distance between the centers of mass of two blobs.
Blobs further interact with colloids or the flat surface at $z=0$ via
an exponential repulsion
\begin{equation} 
\beta V_{bc}(z) = 3.20\, e^{-4.17 (z/R_g - 0.50)}\;, 
\end{equation} 
where $z$ is the distance between the blob's center of mass and the
surface (i.e., $z = r - R_c$ for the blob--colloid interaction).  { This
specific form of the interactions and the values of the constants have
been derived in \cite{BarbaraBlobs}, and used previously in
\cite{Fran1, Fran2, FranSuperSelect}. } To each colloid, we attach a
number~$\kappa$ of DNA chains, with attachment modelled using the
following radial harmonic spring potential:
\begin{equation} 
\beta V_{\text{tether}}(r) = \frac{3}{4}\left(\frac{r - R_c}{R_g}\right)^2\;, 
\end{equation} 
where $r$ is the distance between the blob's center of mass and the
center of the colloid.  This form of attachment allows the DNA to move
freely on the colloid surface~\cite{Fran1}.
 
The colloids move on a DNA-coated surface, with each grafted DNA
represented as a blob grafted at a specific point~$\vr^0$ on the
surface.  The grafting is modelled using a harmonic spring potential,
given by
\begin{equation} 
\beta V_{\text{graft}}(r) = \frac{3}{4}\left(\frac{r}{R_g}\right)^2\;, 
\end{equation} 
where $r$~is the distance between the blob's center of mass and the
grafting point, $\vr^0$.  For a model surface with dimensions~$L\times
L$, we randomly choose $N = \sigma L^2$ grafting points with the
following probability distribution:
\begin{equation} 
P(x^0, y^0, z^0) \propto \left(\frac{x^0}{L}\right) \delta(z^0 - 0). 
\end{equation} 
Here, $\sigma$ is the grafting number density per unit area, and $0
\leq x^0, y^0 < L$.  With this distribution of grafting points, there
is a roughly linear grafting density gradient along the~$x$ axis.  In
all cases discussed below, we average all observations over multiple
realisations of the surface grafting points.
 
DNAs on the colloids have a sticky end that is complementary to an
analogous sticky end on DNA on the surface.  We model this binding as
a binary event: a given pair of complementary DNAs is either bound or
unbound.  If bound, the energy of the system shifts by an
amount~$\beta f$, and the centers of mass of the blobs corresponding
to the DNAs interact via a harmonic spring potential.  Hence, each
bond adds to the potential energy of the system a term given by
\begin{equation} 
\beta V_{\text{bond}}(r) = \beta f + 0.534 (r / R_g - 0.730)^2\;, 
\label{eq:Bb}
\end{equation} 
where $r$ is the distance between the centers of mass of the two
blobs.  { The numerical values of the constants in Eq.~\ref{eq:Bb} 
-- corresponding to the self-avoiding walk model for polymers -- were
calculated in \cite{BarbaraBlobs}.} The binding energy $\beta f$ is 
related by a constant shift to the hybridisation free energy of the 
sticky ends in solution at standard conditions, $\beta \Delta G_0$ 
that can be determined experimentally. Concretely,
\begin{equation} 
\beta f = \beta\Delta G_0 + \ln[\rho_0 R_g^3 q_{AB}], 
\end{equation} 
where $\rho_0 = 1\,$M is the standard concentration, and $q_{AB} =
41.15$ is a constant chosen so that free sticky ends in solution
(modelled as point particles bound by~$V_{\text{bond}}(r)$) have the
same binding constant as their experimental counterparts { 
(see Online SI for a detailed derivation of this relation)}.
For colloids with diameter $100\,$nm, we have $f = \Delta G_0 +
11.6\kT$.
 
\subsection{Explicit simulations}
In summary, the final Hamiltonian for a system with colloids centered at 
$\{\vR_a\}$ and DNAs represented as blobs with centers of mass at 
$\{\vr_i\}$ is 
\begin{multline} 
V = 
\sum_{a<b} V_{HS}(|\vR_a - \vR_b| - \sigma) + 
\sum_a V_{HS}(z_a - \sigma/2) +\\ 
\sum_i V_{bc}(z_i) +  
\sum_{a,i} V_{bc}(|\vR_a - \vr_i| - \sigma/2) +\\ 
\sum_{i<j} V_{bb}(|\vr_i - \vr_j|) + 
\sum_a \sum_{i \in a} V_{\text{tether}}(|\vR_a - \vr_i|) +\\ 
\sum_{i\in S} V_{\text{graft}}(|\vr_i - \vr_i^0|) + 
\sideset{}{'}\sum_{i < j} V_{\text{bond}}(|\vr_i - \vr_j|). 
\end{multline} 
In this expression, $i \in a$ denotes the set of DNAs~$i$ tethered to
colloid~$a$, $i \in S$ denotes the set of DNAs~$i$ grafted on the
surface at points $\{\vr_i^0\}$, and the primed sum is over all bound
pairs of DNA chains, $i$~and~$j$.  Respectively, the terms in the
Hamiltonian capture the hard-sphere interaction between colloids,
between colloids and the surface, the repulsion between DNAs and the
surface, the repulsion between DNAs and colloids, the Gaussian
repulsion between DNAs, the tethering of DNAs to particular colloids,
the grafting of DNAs on the surface, and the binding energy between
bound complementary DNAs.

Dynamics of this model is approximated by kinetic Monte Carlo
simulations. Particle displacements with uniformly distributed step
sizes are attempted at a frequency consistent with the free colloid
having diffusion constant $D = \kT / 6 \pi \eta R_c$, where $\eta
\approx 1\,$cP is the viscosity of water.  These moves define the
correspondence between Monte Carlo steps and real time.  Further, DNA
blob displacements are attempted at the same frequency.  Finally,
hybridization is implemented via bond-forming/bond-breaking MC moves
as described in \cite{Fran1}, with a frequency chosen to mimic
realistic DNA hybridisation kinetics. Such model can realistically
capture the stochastic nature of binding events, which dominates the
dynamics of nanocolloids with relatively few ``legs''.
 
\subsection{Implicit modelling of micron-size colloids: Mean field approach}  
For micron-sized DNACCs, a common grafting setup uses short, dsDNA
tethers (modeled as rigid rods) capped with short sticky
ends~\cite{mirjam-jacs,rogers-crocker, lorenzo-bigels}.  We have used
our previously developed self-consistent mean field
theory~\cite{patrick-jcp,stefano-jcp} to map the free energy landscape
experienced by such colloids.  For concreteness, we have chosen to
model tethers that are $20\,$nm long.  The colloid is taken to have
radius $R = 0.5\,\mu$m and be coated with strands of type $\gamma$,
randomly and uniformly, at a density, $\sigma$, equal to $1$~tether
per~$(5\,$nm$)^2$.  The surface is coated with a gradient of $\alpha$
and $\beta$--type strands (please refer to Fig. 3 for details).  
At one end, the density of $\alpha$-type
strands is $2\sigma$, and there are no $\beta$ strands.  The converse
situation holds at the opposite end of the gradient.  The
parameter~$c$ measures progress along this linear gradient.  The
solution binding free energy of $\alpha$ and $\gamma$ sticky ends is
given by $\Delta G^0 - \delta/2$, while that of $\beta$ and $\gamma$
strands is given by $\Delta G^0 + \delta/2$.  No other pairs of sticky
ends have appreciable binding.
 
To compute the colloid-surface interaction free energy at any point 
$c$ on the surface, we first compute the interaction free energy per 
unit area, $f(h')$, of two plates with coats and grafting densities 
corresponding to the value of $c$, as a function of plate-plate 
separation, $h'$.  We then use the Derjaguin approximation to estimate 
the interaction between a spherical colloid and the planar surface, 
$F(h)$, as a function of the distance of closest approach, $h$, i.e. 
\[ 
F(h) = 2\pi R \int_h^\infty \dee h'\, f(h'). 
\] 
Since $F(h)$ varies strongly with $h$, we assume that the particle 
sits at or very close to the height~$h$ that minimizes $F(h)$, which 
is usually equal to the tether length, i.e. $20\,$nm.  The value of 
$F$ at that height is used to estimate the free energy of binding.  As 
long as the curvature of $F(h)$ around this minimum is independent of 
$c$, this estimate differs from the real binding free energy by a 
constant amount, which does not affect the particle's surface 
dynamics. 
 
\begin{figure} 
\centerline{\includegraphics[width=0.48\textwidth]{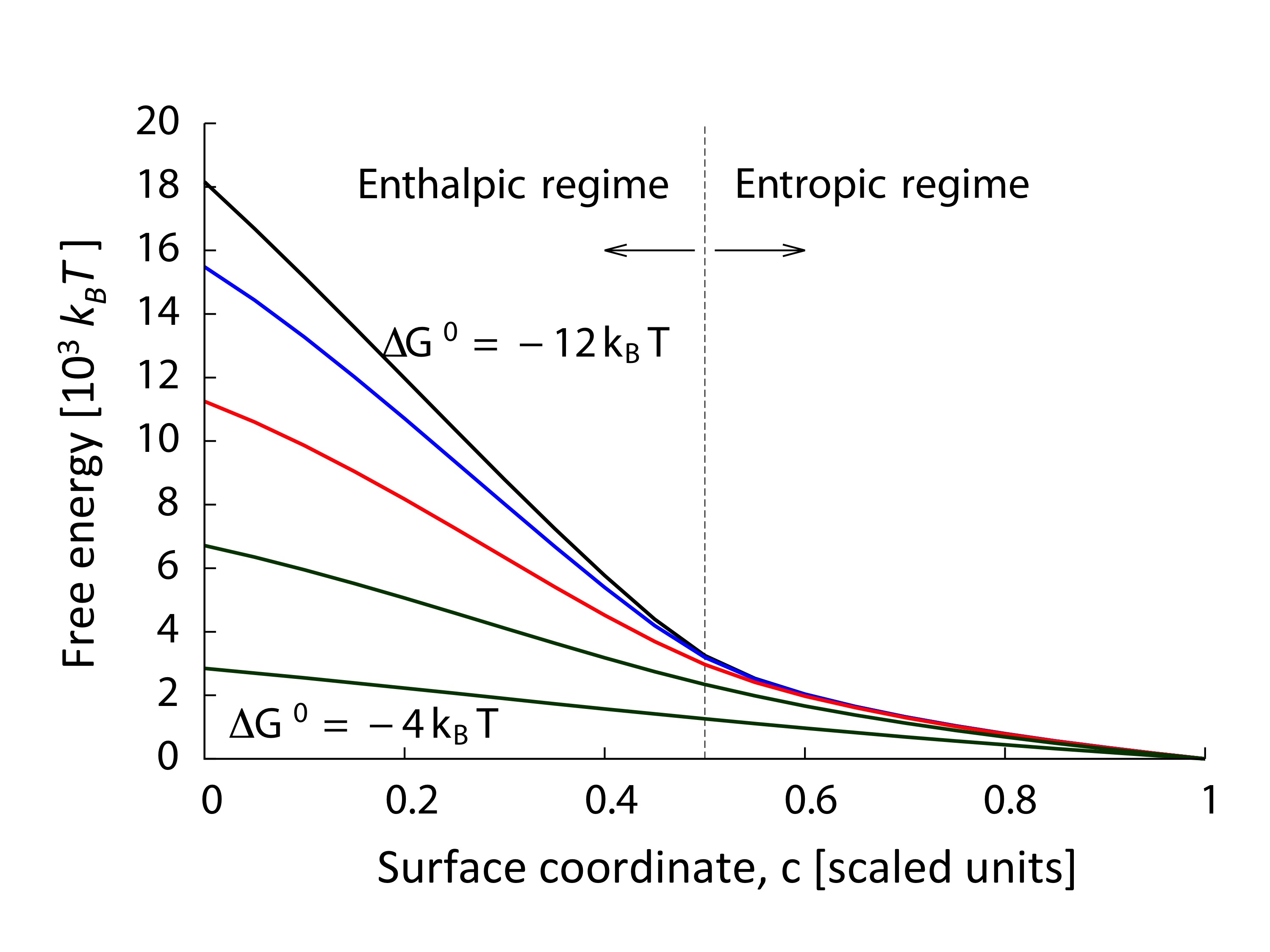}}
\caption{\label{fig:MicronWolloids1}Relative binding free energy of
  micron-sized DNACC to surface along a linear surface coating
  gradient.  The colloid is coated with $\gamma$-type strands with
  density $\sigma = 1/(5\,\text{nm})^2$.  Along the surface, the density
  of $\alpha$-type strands is $2\sigma(1-c)$ and that of $\beta$-type
  strands is $2\sigma c$.  The binding free energy of sticky ends
  $\alpha$ and $\gamma$ in solution is $\Delta G^0 - 4\,$\kT, and that
  of $\beta$ and $\gamma$ strands is $\Delta G^0 + 4\,$\kT.  Free
  energy landscapes are shown for $-\Delta G^0 = 12, 10, 8,
  6$~and~$4\,$\kT. The last curve with $\Delta G^0 = -4\,$\kT\,
  corresponds to $\Delta G_\alpha=0$ and $\Delta G_\beta=-8\,$\kT,
  which is roughly equivalent to the system with only one type of
  receptors.}
\end{figure}

Figure~\ref{fig:MicronWolloids1} shows the free energy of binding as a
function of the progress, $c$, along the surface coating gradient, for
$\delta = 8\,$\kT, as a function of mean sticky-end binding strength,
$\Delta G^0$.  At high overall binding strengths ($\Delta G_0$ around
-12 $\kT$), there are two clearly visible regimes in this free energy
landscape.  For $c < 0.5$, it is impossible for all the $\gamma$-type
strands in the colloid to bind to the more favorable surface
$\alpha$-type strands as opposed to $\beta$-type strands.  Hence,
moving the particle along the surface-coating gradient results in a
decrease in free energy owing to more favorable binding partners being
available to the colloid's tethers.  Essentially all $\gamma$-type
strands on the colloid are bound, and the decrease in free energy is
linear along the gradient, with a slope proportional to $\delta$.  We
call this regime the \emph{enthalpic} regime.  For $c > 0.5$, all
$\gamma$ strands can and do bind to $\alpha$ strands on the surface.
Hence, the gradient in free energy is driven purely by the increase in
the number of $\alpha$ partners available to the colloid.  We call this
regime the \emph{entropic} regime.
 
At low binding strength (i.e., high $\Delta G^0$), entropic effects
play an important role even at $c < 0.5$.  There, many $\gamma$
strands remain unpaired, so that $\alpha$-$\gamma$ bonds are not
saturated at $c = 0.5$.  Indeed, for low enough binding strengths
($\Delta G^0$ larger than about 4 $\kT$), the free energy decreases linearly
with constant slope from one end of the surface coating gradient to
the other end. Decreasing the surface grafting density by an order of
magnitude reduces all binding free energies by about the same amount,
without significantly changing the qualitative features of
Figure~\ref{fig:MicronWolloids1}. A rough estimate of whether this
gradient is experimentally realizable can be obtained by supposing
that the particle will only move appreciably if the free energy for
moving by one tether length ($20\,$nm) is $1\,$\kT.  Under these
conditions, the linear gradient in Figure~\ref{fig:MicronWolloids1}
would have to be no longer than about $100\,\mu$m in size to realize
both the enthalpic and entropic regimes, or no longer than about
$500\,\mu$m to realize only the enthalpic regime.  These sizes should
be within the range of current micro-patterning techniques like
dip-pen nanolithography. The mean field approach presented in this
section has thus proven to be a usefull tool for exploring the rich
phase space of the system as well as to predict novel behaviour.
 
 \section{Conclusions} 
{ We have presented a model study of the diffusion and directed motion
of DNA-coated colloids on DNA-functionalized surfaces.  For the case
of DNACCs coated with a small number of DNA legs, we observed a 
non-monotonic dependence of the diffusion rate on the number of "legs" 
and on their hybridization free energy. As a general consequence of the 
competing kinetic and thermodynamic trends, optimal parameter values 
exist, such that the colloids move the fastest along the receptor 
concentration gradients.}

For the case of DNACCs densely coated with
DNA, we found instead two different diffusion regimes depending on the
single-bond binding energy, and we identify the causes in the
enthalpic-vs-entropic nature of binding in the two different cases.
Finally, we have shown how insights into the thermodynamics of DNA and
DNACCs interactions provide a valuable way to design possible
applications for this system, and we discuss the explicit case of
reversible DNACC walkers as well as the use of DNA-grafted surfaces as
a chromatographic tool to separate DNACCs with different
functionalization, a process that could improve their use in a number
of important technological applications. The underlying mechanisms of
the phenomena that we describe are not specific to particles coated
with DNA.  Any colloidal system with multivalent ligand-receptor
interactions should exhibit similar behaviour.

Thus the approach that we describe above could be used to separate
multivalent nanoparticles according to their valence. Again, such
separations would be potentially very useful for the purification of
multivalent "magic bullets" for targeting specific pathogens
\cite{FranSuperSelect}.

\section{Acknowledgments} 
This work was supported by the 7th Framework Programme of European
Union through grants ARG-ERC-COLSTRUCTION 227758, EPSRC Programme Grant EP/I001352/1 and ITN-COMPLOIDS
234810, and by the Slovenian research agency through Grant P1-0055.
P.V. acknowledges funding from a Marie Curie International Incoming
Fellowship of the European Community FP7 and S.A-U from an Alexander
von Humboldt Postdoctoral fellowship.


\begin{thebibliography}{10} 
 
\bibitem{mirkin} Mirkin, C. A. and Letsinger, R. C. and Mucic, 
  R. C. and Storhoff, J. J. {\em A {DNA}-based method for rationally 
    assembling nanoparticles into macroscopic materials} Nature 382 
  (1996) pp.~607--609 
 
\bibitem{alivisatos} Alivisatos, A. P. and Johnsson, K. P. and Peng, 
  X. and Wilson, T. E. and Loweth, C. J. and Bruchez, M. P. and 
  Schultz, P. G. {\em Organization of 'nanocrystal molecules' using 
    {DNA}} Nature 382 (1996) pp.~609--611
 
\bibitem{mirkin-crystal} Park, S. Y, Lytton-Jean, A. K. R., Lee, B., 
  Weigand, S., Schatz, G. C. and Mirkin, C. A., {\em 
    {DNA}-programmable nanoparticle crystallization} Nature Materials 
  451 (2008) pp 553-556 
 
\bibitem{gang-crystal} Nykypanchuk, D., Maye, M. M., van der Lelie, 
  D. and Gang, O., {\em {DNA}-guided crystallization of colloidal 
    nanoparticles} Nature Materials 451 (2008) pp.~549-552 

\bibitem{crocker-tweezers} Biancaniello, P. L. and Kim, A. J. and 
  Crocker, J. C., {\em Colloidal Interactions and Self-Assembly Using 
    {DNA} Hybridization}, Physical Review Letters 94 (2005) pp.~058302 
 
\bibitem{rogers-crocker} Benjamin, R.~W.  and Crocker, J.~C., 
  {\em Direct measurements of DNA-mediated colloidal interactions and 
    their quantitative modeling} Proceedings of the National Academy 
  of Sciences 108 (2011) pp.~15687-15692 

\bibitem{control-position} Suzuki, K., Hosokawa, K. and Maeda, M.,
  {\em Controlling the Number and Positions of Oligonucleotides on
    Gold Nanoparticle Surfaces}, Journal of the American Chemical
  Society 131 (2009), pp.~7518--7519
 
\bibitem{mirjam-nature} Leunissen, M. E., Dreyfus, R., Cheong, F. C., 
  Grier, D.G., Sha, R., Seeman, N. C. and Chaikin, P. M., {\em 
    Switchable self-protected attractions in {DNA}-functionalized 
    colloids} Nature Materials 8 (2009) pp.~590-595 
 
\bibitem{mirjam-jacs} Leunissen, M. E., Dreyfus, R., Sha, R., Seeman, 
  N. C. and Chaikin, P. M., {\em Quantitative Study of the Association 
    Thermodynamics and Kinetics of {DNA}-Coated Particles for 
    Different Functionalization Schemes} Journal of the American 
  Chemical Society 132 (2010) pp.~1903-1913 
 
\bibitem{dreyfus} Dreyfus, R and Leunissen, M E and Sha, R and 
  Tkachenko, A and Seeman, N C and Pine, D J and Chaikin, P M {\em 
    Aggregation-disaggregation transition of DNA-coated colloids: 
    Experiments and theory}, Physical Review E 81 (2010) pp.~41404 
 
\bibitem{lorenzo-bigels} Varrato, F., Di Michele, L., Belushkin,
  M. ,Dorsaz, N. Nathan, S.H., Eiser,E. and Foffi, G.  {\em Arrested
    demixing opens route to bigels} Proceedings of the National
  Academy of Sciences USA 109 (2012) pp.~19155-19160

\bibitem{mirkin-crystal2} Macfarlane, R.~J., Jones, M.~R., Lee, B.,
  Auyeung, E. and Mirkin, C.A.  {\em Topotactic Interconversion of
    Nanoparticle Superlattices} Science 341 (2013) pp.~1222--1225

\bibitem{gang-crystal2} Zhang, Y., Lu, F., Yager, K.~G., van der
  Lelie, D. and Oleg Gang {\em A general strategy for the DNA-mediated
    self-assembly of functional nanoparticles into heterogeneous
    systems} Nature Nanotechnology (2013) Advanced Online Publication,
  October 2013

\bibitem{lorenzo-kinetics} Di Michele, L., Varrato, F.,Kotar, J.,
  Nathan, S.H., Foffi,G. \& Eiser, E.  {\em Multistep kinetic
    self-assembly of DNA-coated colloids} Nature communications 4
  (2013)

\bibitem{stefano-nature} Angioletti-Uberti, S., Mognetti, B.M. and 
  Frenkel, D., {\em Reentrant melting as a design principle for 
    DNA-coated colloids} Nature Materials 11 (2012) pp.~518-522 
 
\bibitem{tkachenko} Tkachenko, A.~V. {\em Morphological Diversity 
  of DNA-Colloidal Self-Assembly} Physical Review Letter 89 (2002) 
  148303 
 
\bibitem{bortolo-pnas} Mognetti, B. M., Varilly, P., 
  Angioletti-Uberti, S., Martinez-Veracoechea, F. J., Dobnikar, J., 
  Leunissen, M. E. and Frenkel, D. {\em Predicting DNA-mediated 
    colloidal pair interactions} Proceedings of the National Academy 
  of Sciences (2012), online edition only 
 
\bibitem{patrick-jcp} Varilly, P., Angioletti-Uberti, S., Mognetti, 
  B.M. and Frenkel, D., {\em A general theory for DNA-mediated and 
    other valence limited interactions} Journal of Chemical Physics 
  137 (2012) pp.~094108-094123
   
\bibitem{stefano-jcp} Angioletti-Uberti, S.,Varilly, P., Mognetti,
  B.M., Tkachenko, A.V. and Frenkel, D., {\em Communication: a simple
    analytical formula for the free-energy of ligand-receptor mediated
    interactions} Journal of Chemical Physics 138 (2013)
  pp.~21102-21106
   
 \bibitem{bortolo-competing} Mognetti, B.~M., Leunissen, M.~E. and 
  Frenkel, D. {\em Controlling the temperature sensitivity of 
    {DNA}-mediated colloidal interactions through competing linkages} 
  Soft Matter 8 (2012) pp.~2213, 
   
\bibitem{olvera} Li, T.I.N.G, Sknepnek,R.  Macfarlane,R.J., Mirkin,
  C.A and Olvera de la Cruz, M. {\em Modeling the Crystallization of
    Spherical Nucleic Acid Nanoparticle Conjugates with Molecular
    Dynamics Simulations} Nano Letters 12 (2012) pp.~2509-2514
 
\bibitem{travesset} Knorowski C, Burleigh S and Travesset A .  {\em
  Dynamics and Statics of DNA-Programmable Nanoparticle Self-Assembly
  and Crystallization} Physical Review Letters. 106 (2011) pp.~215501
   
\bibitem{Fran2} F.~J. Martinez-Veracoechea, B. M. Mladek, 
  A.~V. Tkachenko, and D. Frenkel, {\em Design Rule for Colloidal 
    Crystals of DNA-Functionalized Particles} Phys. Rev. Lett. 107 
  (2011) pp.~045902-1--045902-4 
 
\bibitem{bianca} Mladek, B.~M., Fornleitner, J., Martinez-Veracoechea,
  F.~J., Alexandre, D. and Frenkel, D. \em{Quantitative Prediction of
    the Phase Diagram of DNA-Functionalized Nanosized Colloids}
  Physical Review Letters 108 (2012) pp.~268301-268305
 
\bibitem{lorenzo-review} Di Michele, L. and Eiser, E.  {\em
  Developments in understanding and controlling self assembly of
  DNA-functionalized colloids} Physical Chemistry Chemical Physics 15
  (2013) pp.~ 3115-3129
 
\bibitem{travesset-review} Knorowski C and Travesset A .,
{\em Materials design by DNA programmed self-assembly},
 Current Opinion in Solid State \& Materials Science. 15 (2011) pp.~262-270

\bibitem{turberfield-walker} Yurke, B. \emph{et. al.} {\em A 
  DNA-fuelled molecular machine made of DNA} Nature 406 (2000) 
  pp.~605--608 
  
\bibitem{seeman-walker} Sherman, W.~B. and Seeman, N.C. 
       {\em A Precisely Controlled DNA Biped Walking Device} Nano 
       Letters 4 (2004) pp.~1203-1207 
 
 \bibitem{turberfield-synthesis} McKee, M.~L. and Milnes, P.~J. and
   Bath, J., Stulz, E., O'Reilly, R.~K. and Turberfield, A.~J., {\em
     Programmable One-Pot Multistep Organic Synthesis Using DNA
     Junctions}, Journal of the American Chemical Society 134 (2012)
   pp.~1446-1449
     
\bibitem{seeman2} Gu, H., Chao, J., Xiao, S.~J.  and Seeman,N.~C.  {\em A 
  Proximity-Based Programmable DNA Nanoscale Assembly Line}, Nature 
  465 (2010) pp.~202-205 
  
\bibitem{liu} He, Y. and Liu, D.~R.  {\em Autonomous multistep organic
  synthesis in a single isothermal solution mediated by a DNA walker}
  Nature Nanotechnology 5 (2010), pp.~778Ð782

\bibitem{dna-computing1} Stojanovic, M.~N., Mitchell, T.~E. and
  Stefanovic, D.  {\em Deoxyribozyme-Based Logic Gates} Journal of the
  American Chemical Society 124 (2002), pp.~3555-3561

\bibitem{dna-computing2} Qian, L. and Winfree, E.  {\em Scaling Up
  Digital Circuit Computation with DNA Strand Displacement Cascades}
  Science 332 (2011), pp.~1196-1201
  
\bibitem{Fran1} F.~J. Martinez-Veracoechea, B. Bozorgui, and 
  D. Frenkel, {\em Anomalous phase behavior of liquid-vapor 
    phase transition in binary mixtures of DNA-coated particles} Soft 
  Matter 6 (2010) pp.~6136--6145 
 
\bibitem{BarbaraBlobs} Pierleoni, C., Capone, B. and Hansen,J.~P.  {\em 
  A soft effective segment representation of semidilute polymer 
  solutions} J. Chem. Phys. 127 (2007) pp.~171102 
 
\bibitem{Crocker_Tau} Rogers, W.B., Sinno, T. and Crocker, J.~C. {\em
  Kinetics and non-exponential binding of DNA-coated colloids}, Soft
  Matter {\bf 9} (2013) pp.~6412--6417

\bibitem{santalucia} SantaLucia, J.  {\em{A unified view of polymer,
    dumbbell, and oligonucleotide {DNA}
    nearest-neighborâthermodynamics}} Proceedings of the
  National Academy of Sciences of the United States of America 95
  (1998) pp.~1460--1465

\bibitem{DINAMelt} Markham, N.~R.,  Zuker, M. {\em DINAMelt web 
server for nucleic acid melting prediction}, Nucleic Acids Res. 
{\bf 33} (2005) pp.~W577-W581

\bibitem{tkachenko-pre} Licata, N.~A., Tkachenko, A.~V. {\em Colloids
  with key-lock interactions: Nonexponential relaxation, aging, and
  anomalous diffusion}, Phys. Rev. E {\bf 76} (2007) pp.~041405

\bibitem{vitrimers} Montarnal, D., Capelot, M., Tournilhac, F., Leibler, L. 
{\em Silica-Like Malleable Materials from Permanent Organic Networks}, 
Science {\bf 334} pp.~965--968

\bibitem{Winfree} Zhang, D.~Y., Winfree, E. {\em Control of DNA 
strand displacement kinetics using toehold exchange},  
J. Am. Chem. Soc. {\bf 131} (2009) pp.~17303--17314

\bibitem{FranSuperSelect} Martinez-Veracoechea, F.~J. and Frenkel, D. 
  {\em Designing super selectivity in multivalent nano-particle 
    binding} Proceedings of the National Academy of Sciences 108 
  (2011) pp.~10963--10968 


\end{thebibliography}
\end{document}